\documentclass[prb,twocolumn]{revtex4-1}

\usepackage{amssymb}
\usepackage{amsmath}
\usepackage[dvips]{graphicx}

\begin{document}

\title{Magnetostatic fields in tubular nanostructures}
\author{P. Landeros$^{1}$, P. R. Guzm\'{a}n$^{2}$, R. Soto-Garrido$^{3}$,
and J. Escrig$^{4,5}$}

\address{$^1$
Departamento de F\'{i}sica, Universidad T\'{e}cnica Federico Santa
Mar\'{i}a, Avenida Espa\~{n}a 1680, Casilla 110 V, 2340000
Valpara\'{i}so, Chile} 
\address{$^2$
Departamento de Matem\'{a}tica, Universidad T\'{e}cnica Federico
Santa Mar\'{i}a, Avenida Espa\~{n}a 1680, Casilla 110 V, 2340000
Valpara\'{i}so, Chile} 
\address{$^3$ Department of Physics, University of
Illinois at Urbana-Champaign, 1110 W. Green Street, Urbana, 61801
Illinois, USA} 
\address{$^4$ Departamento de F\'{i}sica,
Universidad de Santiago de Chile (USACH), Avenida Ecuador 3493,
917-0124 Santiago, Chile} 
\address{$^5$
Centro para el Desarrollo de la Nanociencia y Nanotecnolog\'{i}a,
CEDENNA, 917-0124 Santiago, Chile.}

\begin{abstract}
The non-uniform magnetostatic field produced by the equilibrium and non
equilibrium magnetic states of magnetic nanotubes has been investigated
theoretically. We consider magnetic fields produced by actual equilibrium
states and transverse and vortex domain walls confined within the
nanostructure. Our calculations allow us to understand the importance of the
magnetostatic field in nanomagnetism, which is frequently considered as a
uniform field. Moreover, our results can be used as a basis for future
research of others properties, such as the investigation of spin waves when
domain walls are present, or the motion of a magnetic particle near a
magnetic field.
\end{abstract}

\maketitle

\section{Introduction}

Structures of nanometric dimensions are strongly dependent of size and shape
mainly due to the fact that the characteristic scales of many physical
phenomena are comparable to the dimensions of current nanostructures.
Magnetism makes no exception and fundamental research of nanomagnets is
further fuelled by their prospective applications into technological devices 
\cite{WAB+01,GBH+02} and biomedical applications \cite{ET03,Lee07}. The
implementation of nanomagnets into potential devices requires a deep control
of their physical properties; hence it is very important to control the
shape and size of the nanoparticles which are crucial to determine their
magnetic properties. Within currently available nanostructures, nanotubes
made from 3d transition metals and their alloys are of particular
importance, since the nanotube topologies provide us of two surfaces for
modification, allowing the generation of multifunctional magnetic
nanoparticles. Clearly, a technology capable to modify different surfaces
would be highly desirable \cite{Eisenstein05}.

More recently hollow tubular nanostructures have been synthesized \cite%
{SRH+05,NCM+05,WLL+05,TGJ+06} and may be useful for applications in
biotechnology, because low density magnetic nanotubes (MNs) can float in
solutions and are more suitable for in vivo applications \cite{Eisenstein05}%
. In recent years, the magnetism of planar and circular nanowires has been
intensely investigated, while MNs have received less attention, although MNs
do not exhibit magnetic vortex cores or Bloch points \cite{Hertel04},
leading to a different behaviour with a more controllable reversal process.

It is well know that changes in the cross section of the MNs strongly affect
the reversal mechanism and the overall magnetic behaviour \cite%
{WLL+05,DKG+07,ELA+07-1,ELA+07-2,LSC+09}. Recent articles have shown that
MNs present three main equilibrium states: a uniform state (U), a mixed
state (M) and a vortex state (V). With basis in theoretical models it has
been argued that for nanotubes with outer radius less than $4L_{x}$ (where $%
L_{x}=\sqrt{2A/\mu _{0}M_{s}^{2}}$ is the exchange length, $M_{s}$ the
saturation magnetization and $A$ the stiffness constant of the magnetic
material) the magnetization is almost uniform and oriented principally along
the cylindrical axis \cite{LSC+09}. We call this configuration uniform
state. Also for nanotubes with $R\gtrsim 4L_{x}$ the magnetic equilibrium
state correspond to the so-called mixed state, which is a mixture of U and V
states \cite{WLL+05,LSC+09,LSS+07,CUB+07}. This state presents a uniform
magnetization along the middle region of the tube and near the lower and
upper surfaces the magnetization deviates from the uniformity in order to
reduce the magnetostatic field \cite{LSC+09}. Finally, there is a transition
from the M state to the vortex state as can be seen from figure 7 of \cite%
{LSC+09}.

On the other hand, it has been argued that domain wall propagation in
nanostructures, which can be achieved by external fields or spin-currents,
is of basic scientific and technological interest \cite{AAX+03, THJ+06}.
More recently, it has been shown that the size-dependent reversal process in
MNs occurs via DW nucleation and subsequent DW propagation. The DW
microstructure depends on the tube cross section and can be a transverse
wall (TW) for tube radius ($R$) smaller than a critical radius ($R_{c}(\beta
)$), or a vortex wall (VW) for $R>R_{c}(\beta )$ \cite%
{LAE+07,EBJ+08,AEA+08,BEP+09}, where $\beta\equiv R_{i}/R$ is the ratio
between the inner ($R_{i}$) and outer tube radius. The critical radius
defined above depends on the magnetic material and on the shape factor $%
\beta $ and then it can be tailored manipulating its geometrical parameters.
In polycrystalline systems, the critical radius, $R_{c}(\beta)$, ranges from
a few nanometers to 20 nm approximately, and therefore, since experimentally
fabricated MNs satisfy $R>R_{c}(\beta)$, we can expect that the propagation
of a vortex DW be the dominant magnetization reversal mechanism for
ferromagnetic nanotubes \cite{LAE+07,EBJ+08,AEA+08,BEP+09}.

Although the magnetostatic field of MNs with uniform magnetization has been
already investigated \cite{EAA+08,PDE09,EAA+09}, an analytical model for the
magnetostatic field produced by a MN with the above mentioned
configurations, either equilibrium or transient states, has not yet been
performed. Our intention is to fill this gap by using simple magnetization
models \cite{LSC+09, LAE+07}. Since nanostructures are usually
polycrystalline, the crystallographic orientations of the crystallites are
random and, as a consequence, the average magnetic anisotropy of the
particle is very small. In view of that, it will be neglected in our
calculations \cite{KVL+03, CRE+00}. This paper is organized as follows: In
Sec. \ref{Model} we present the magnetization models, including in Sec. \ref%
{ModelA} the equilibrium states M and U and the VW, and then in Sec. \ref%
{ModelB} we discuss the case of the TW. Results and discussions are
summarized in Sec. \ref{Results} and conclusions in Sec. \ref{Remarks}.

\section{Calculation of the magnetostatic fields}

\label{Model}

In this section we present explicit expressions to evaluate the
magnetostatic field for nanotubes with the magnetization states described
above. Within the continuum theory of ferromagnetism \cite{Aharoni96}, the
magnetostatic field is given by $\mathbf{H}=-\nabla U$, where $U$ is the
magnetostatic potential, which depends on the magnetization itself. The
basic assumption we address in this paper is that the nanotubes we are
considering are thin enough to avoid any dependence of the magnetization in
the radial coordinate. In what follows we present the magnetization and the
related magnetostatic field for the equilibrium states (U and M) and both
DWs structures aforementioned.

\subsection{Equilibrium states and vortex-like configurations}

\label{ModelA}

First, we consider nanotubes with vortex-like configurations, as the mixed
state and the transient state with a VW confined within the tube. The
magnetization of such magnetic configurations can be generally written as $%
\mathbf{M}(z)=M_{s}\mathbf{m}(z)=M_{s}(m_{\phi }(z)\hat{\phi }+m_{z}(z)\hat{%
\mathbf{z}})$, where $m_{z}(z)=\cos \Theta (z)$, and $m_{\phi }(z)=\sin
\Theta (z)$, with $\Theta (z)$ being the angle between the local
magnetization vector and the cylindrical $z$ axis. With this rather general
form for the magnetization we can compute the magnetostatic field produced
by $\mathbf{M}(z)$. Following the calculations of \cite{LSC+09}, the
magnetostatic potential can be written as%
\begin{equation}
U(\rho ,z)=\frac{M_{s}}{2}\int\nolimits_{0}^{\infty }g(q)J_{0}(q\rho )\left[
f_{s}+f_{v}\right] dq,
\end{equation}%
where $f_{s}$ and $f_{v}$ come from the surface and volumetric magnetic
charges and can be written as 
\begin{equation}
f_{s}\equiv -m_{z}(0)e^{-q|z|}+m_{z}(L)e^{-q|z-L|},
\end{equation}%
\begin{equation}
f_{v}\equiv -\int_{0}^{L}\frac{\partial m_{z}(y)}{\partial y}e^{-q|z-y|}dy.
\end{equation}%
In the potential we have also defined: 
\begin{equation}
g(q)\equiv (R/q)\left[ J_{1}(qR)-\beta J_{1}(\beta qR)\right] ,
\end{equation}%
where $J_{m}(x)$ are Bessel functions. The magnetostatic field associated
with the above potential can be written as $\mathbf{H}=-\nabla U(\rho ,z)$,
and then, the cylindrical components of the field are $H_{\phi }=0$, 
\begin{equation}
H_{\rho }=\frac{M_{s}}{2}\int\nolimits_{0}^{\infty }qg(q)J_{1}(q\rho )\left[
f_{s}+f_{v}\right] dq,
\end{equation}%
\begin{equation}
H_{z}=-\frac{M_{s}}{2}\int\nolimits_{0}^{\infty }g(q)J_{0}(q\rho )\left[ 
\frac{\partial f_{s}}{\partial z}+\frac{\partial f_{v}}{\partial z}\right]
dq.
\end{equation}%
At this point we are in position to calculate the magnetostatic field
through the reversal process for MNs, as well as to investigate the changes
in the magnetostatic field when the equilibrium state is properly
considered. In what follows we will use the equations above to calculate the
magnetostatic field for the M state and the vortex DW.

\subsubsection{Equilibrium magnetization states}

The arrangement of magnetic moments in a MN with small radius can be seen as
a uniform state in the middle region of the tube, but with vortex-like
deviations of the magnetization located at the tube ends \cite%
{WLL+05,LSC+09,LSS+07,CUB+07}. This mixed state has been observed by
micromagnetic simulations \cite{WLL+05,LSS+07,CUB+07} and studied
theoretically \cite{LSC+09}. It has been shown that two equilibrium states
can be studied with the model for the M state. In other words, the U state
is a special case of the former, as we will see later. The model we use for
the M state is given by the function \cite{LSC+09} 
\begin{equation}
\Theta (z)=\left\{ 
\begin{array}{c}
\theta _{0}\left( d-z\right) /d\text{ \ \ \ \ \ \ \ \ \ for }0\leq z\leq d
\\ 
0\text{ \ \ \ \ \ \ \ \ \ \ \ \ \ \ \ \ \ \ for }d\leq z\leq L-\lambda \\ 
\theta _{L}\left( z-L+\lambda \right) /\lambda \text{ \ \ \ \ for }L-\lambda
\leq z\leq L%
\end{array}%
\right.
\end{equation}%
where $d$ and $\lambda $ are the dimensions of the regions where the
magnetization deviates from the $z$-axis. In the model, $\theta _{0}$ and $%
\theta _{L}$ correspond to angles between the magnetization and the $z$-axis
evaluated at the tube ends. Note that we obtain the uniform magnetization
state in the limit $m_{z}(0)=m_{z}(L)=1$ ($\theta _{0}=\theta _{L}=0$) which
occurs approximately for a radius smaller than 4$L_{x}$ \cite{LSC+09}. The
parameters of the model are such that they minimize the total energy for
each set of geometrical parameters. For simplicity we consider symmetric
nanotubes, that is, tubes without difference between both tube ends. In
those cases the four parameters ($d$, $\lambda $, $\theta _{0}$, $\theta
_{L} $) variational problem is reduced to a two parameters variational
problem, when $d=\lambda $ and $\theta _{0}=\theta _{L}$ \cite{LSC+09}.
Then, $f_{s}$ and $f_{v}$ become 
\begin{equation}
f_{s}^{M}=\cos \theta _{0}\left( -e^{-q|z|}+e^{-q|z-L|}\right) ,
\end{equation}%
\begin{equation}
f_{v}^{M}=\frac{\theta _{0}}{d}\int_{0}^{d}\sin (\theta _{0}\frac{d-y}{d}%
)[e^{-q|z+y-L|}-e^{-q|z-y|}]dy.
\end{equation}%
With these quantities in hand we can compute the magnetostatic field $%
\mathbf{H}(\rho ,z)$ at any point. Also, the uniform magnetization state is
recovered with $\theta _{0}=0$, where the functions reduces to $%
f_{s}^{U}=-e^{-q|z|}+e^{-q|z-L|},$ $f_{v}^{U}=0$.

\subsubsection{Vortex domain wall}

The magnetization of the vortex wall in a nanotube can be expressed in a
similar way to the magnetization of the mixed state, that is $\mathbf{M}%
(z)=M_{s}\sin \Theta (z)\hat{\phi }+M_{s}\cos \Theta (z)\hat{\mathbf{z}}$,
but now 
\begin{equation}
\Theta ^{VW}(z)=\left\{ 
\begin{array}{c}
0\text{ \ \ \ \ \ \ \ \ \ \ \ \ \ \ \ \ \ \ \ \ \ \ \ \ \ \ for }0\leq z\leq
z_{w}-\frac{\omega }{2} \\ 
\frac{\pi }{\omega }\left( z-z_{w}\right) +\frac{\pi }{2}\text{\ \ \ \ \ \
for }z_{w}-\frac{\omega }{2}\leq z\leq z_{w}+\frac{\omega }{2} \\ 
\pi \text{\ \ \ \ \ \ \ \ \ \ \ \ \ \ \ \ \ \ \ \ \ \ \ \ \ \ for }z_{w}+%
\frac{\omega }{2}\leq z\leq L%
\end{array}%
\right.
\end{equation}%
as presented in \cite{LAE+07}. Here $z_{w}$ is the location of the center of
the vortex wall, and $w$ is the wall width. Within the above model, the
magnetostatic field can be also expressed with equations (7) and (8), but
now the functions $f_{s}$ and $f_{v}$\ become%
\begin{equation}
f_{s}^{VW}=-e^{-q|z|}-e^{-q|z-L|},
\end{equation}%
\begin{equation}
f_{v}^{VW}=\frac{\pi }{w}\int_{z_{w}-\frac{w}{2}}^{z_{w}+\frac{w}{2}}\cos
\left( \pi \frac{y-z_{w}}{w}\right) e^{-q|z-y|}dy.
\end{equation}

\subsection{Transverse domain wall}

\label{ModelB}

The magnetization of the transverse wall in a nanotube can be expressed in a
similar way to the vortex wall, but with $\mathbf{M}(z)=M_{s}\sin \Theta (z)%
\hat{\mathbf{x}}+M_{s}\cos \Theta (z)\hat{\mathbf{z}}$ \cite{LAE+07}, and $%
\Theta (z)$ being the same of the vortex wall. It can be noted that the main
difference between vortex and transverse walls (in nanotubes) is the
direction of the in-plane magnetization, which makes the nature of the wall
properties to be different. The model that we use for the transverse wall 
\cite{LAE+07} leads us to understand some basic properties, like the DW
width or the energy barrier for the reversal process. Here we are interested
in the visualization of the magnetostatic field produced by a static domain
wall. It can be shown that this field can be written as $\mathbf{H}=-\nabla
U(\rho ,\phi ,z)$, without the usual axial symmetry of vortex-like states.
The case of a vortex wall presents a magnetostatic potential independent of
the angular coordinate $\phi $, which secures angular symmetry. Following 
\cite{LAE+07}, the field can be written as $\mathbf{H}^{TW}=H_{\rho }^{TW}%
\hat{\rho }+H_{\phi }^{TW}\hat{\phi }+H_{z}^{TW}\hat{\mathbf{z}}$. Provided
the magnetostatic potential of the TW can be written as 
\begin{equation}
U^{TW}=U^{VW}+A(\rho ,z)\cos \phi ,
\end{equation}%
the components of the dipole field created for the TW can then be expressed
as 
\begin{equation}
H_{\rho }^{TW}(\rho ,\phi ,z)=H_{\rho }^{VW}(\rho ,z)+\frac{\partial A(\rho
,z)}{\partial \rho }\cos \phi ,  \label{HTrho}
\end{equation}%
\begin{equation}
H_{\phi }^{TW}(\rho ,\phi ,z)=-A(\rho ,z)\sin \phi ,  \label{HTphi}
\end{equation}%
\begin{equation}
H_{z}^{TW}(\rho ,\phi ,z)=H_{z}^{VW}(\rho ,z)+\frac{\partial A(\rho ,z)}{%
\partial z}\cos \phi ,  \label{HTz}
\end{equation}%
where 
\begin{equation}
A(\rho ,z)=\frac{M_{s}}{2}\int\nolimits_{0}^{\infty }qg(q)J_{1}(q\rho
)\varsigma (z,q)dq,  \label{A}
\end{equation}%
and%
\begin{equation}
\varsigma (z,q)\equiv \int_{0}^{L}m_{x}(z^{\prime })e^{-q|z-z^{\prime
}|}dz^{\prime }.
\end{equation}%
Here, $m_{x}(z)=\sin \Theta ^{TW}(z)$, with $\Theta ^{TW}(z)=\Theta ^{VW}(z)$%
. The function $\varsigma (z,q)$ possesses the information about the DW
microstructure and can be straightforwardly integrated. Finally, notice that
the widths of the vortex and transverse walls are essentially different, and
should be evaluated within the framework presented in \cite{LAE+07}.

\section{Results and discussion}

\label{Results}

In the above sections we have obtained explicit expressions to calculate the
magnetostatic field for different magnetic states in nanotubes. In what
follows we present plots of the magnetostatic field produced by the magnetic
states relevant to nanotubes. We begin the discussion with the equilibrium
states (U and M) and then we analyze the case of vortex and transverse walls
propagating along the nanostructure.

\subsection{Vector plots of the equilibrium states}

As mentioned earlier, the uniform magnetization state ($\theta _{0}=0$) is a
special case of the mixed state ($\theta _{0}>0$), and occurs when the tube
radius is approximately less than 4$L_{x}$, provided the exchange energy
dominates over the dipolar one \cite{LSC+09}. When $R>4L_{x}$\ the
magnetization at the tube ends is not entirely parallel to the cylindrical $z
$ axis, instead it develops vortex domains at the ends in order to reduce
the dipolar energy \cite{LSC+09}. The long range character of the dipolar
interaction makes that the dipolar energy grows as $N^{2}$, whereas the
exchange energy grows as $N$, with $N$ the number of atomic magnetic moments
in the tube. Therefore, the edge vortex domains increases in size with the
radius. In the language of the above equations, the size of the vortex
domains is given by the parameters $\theta _{0}$\ and $d$. As $R$ increases
over 4$L_{x}$, $\theta _{0}$ and $d$\ increases, as the reader can see from
figures 5 and 6 in \cite{LSC+09}. To calculate the magnetostatic field we
need to know the set of parameters ($\theta _{0},d$) which minimize the
total energy of the mixed state. Using the expressions for the total energy
of \cite{LSC+09}, we obtain for nanotubes with $\beta =0.9$ and $L=1000L_{x}$
different values of ($\theta _{0},d$) depending on their radius. These
results are summarized in table \ref{table1}. Note that for $L=1000L_{x}$
the transition to a flux-closure vortex state occurs approximately at $%
R\approx 27L_{x}$ \cite{LSC+09}.

\begin{equation*}
\begin{tabular}{rrr}
$R/L_{x}$ & $\theta _{0}$ & $d/L_{x}$ \\ 
$4$ & $0$ &  \\ 
$6$ & $48.7$ & $15.5$ \\ 
$10$ & $75.2$ & $41$ \\ 
$20$ & $83.3$ & $202.6$ \\ 
$27$ & $83.8$ & $414.3$%
\end{tabular}%
\end{equation*}

($\theta _{0},d$) for magnetic nanotubes with $\beta =0.9$\ and $L=1000L_{x}$
as a function of their radius.

\begin{figure}[h]
\begin{center}
\includegraphics[width=8cm]{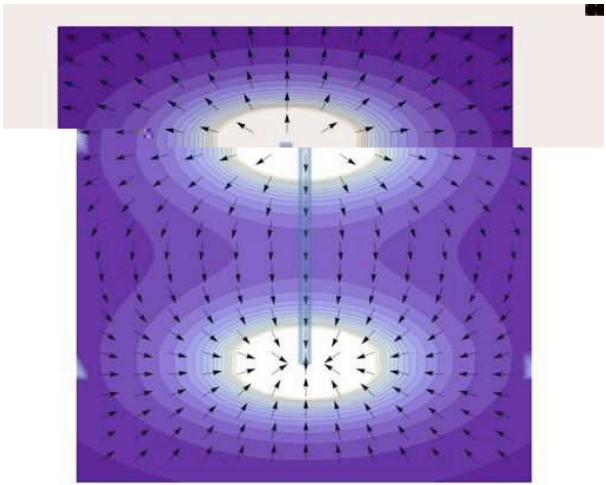}
\end{center}
\caption{(Colour online). Magnetostatic field produced by a tube with
uniform magnetization along the symmetry axis. The arrows have been
normalized to the local strength of the field indicated by the colour.
Parameters are indicated in the text.}
\label{Ferro}
\end{figure}
\begin{figure}[h]
\begin{center}
\includegraphics[width=8cm]{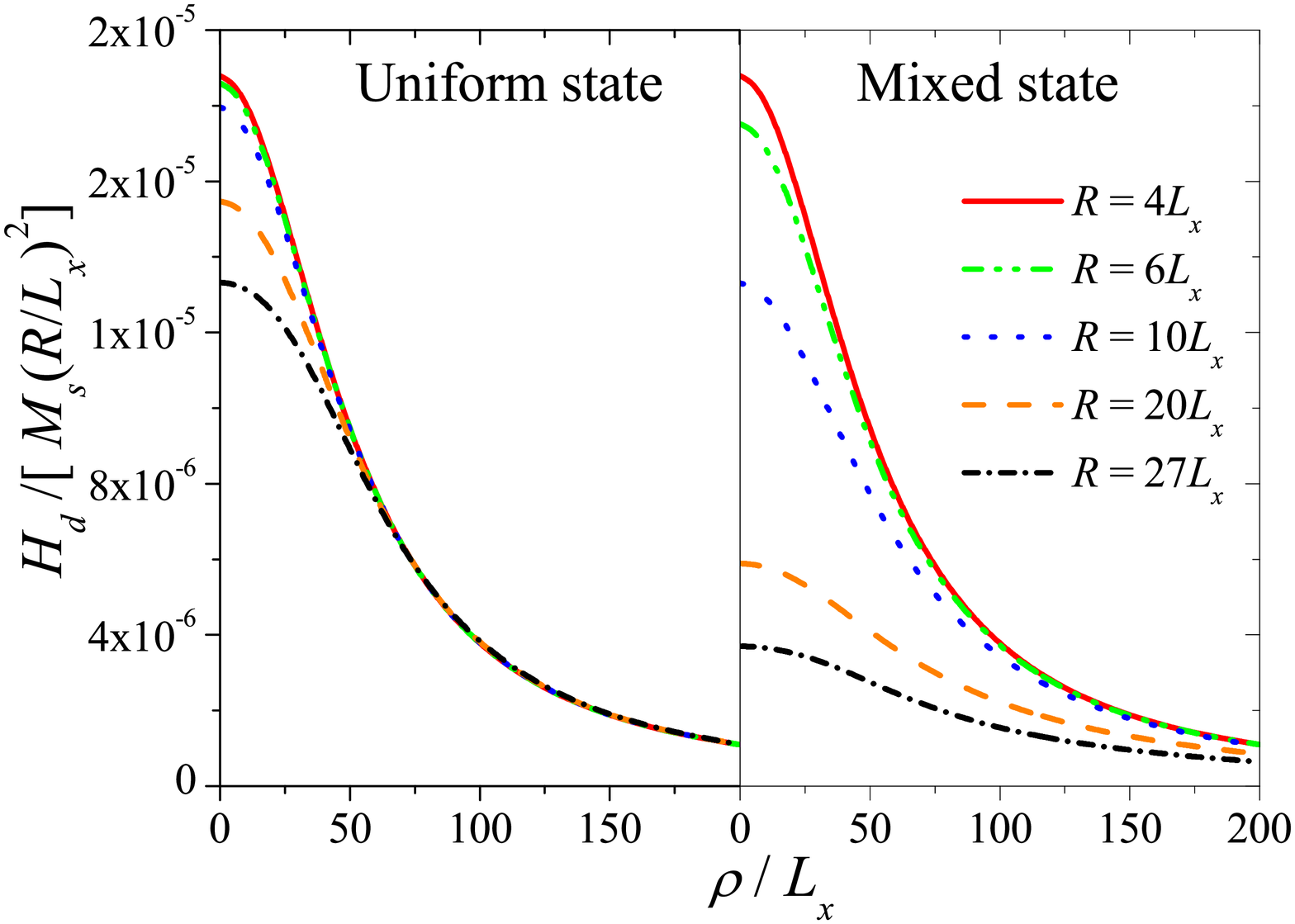}
\end{center}
\caption{(Colour online). Magnetostatic field strength as a function of the
radial coordinate and $z=1050L_{x}$. The left panel shows curves for the
uniform state and the right panel shows the magnetostatic field for the
mixed state. Different curves represent different tube radius, as depicted
in the figure, and parameters have been taken from table 1. }
\label{Hdvsrho}
\end{figure}

The arrows in figure \ref{Ferro} show the normalized magnetostatic field for
a uniformly magnetized MN with $R=4L_{x}$. The colour code indicates the
strength of the field at each point; white colour indicates that the
magnetostatic field is stronger at the tube ends because \textquotedblleft
magnetic charges\textquotedblright\ ($\sigma _{M}$) resides only at the
surfaces such that $\sigma _{M}=\mathbf{M}\cdot \hat{\mathbf{n}}\neq 0$,
with $\hat{\mathbf{n}}$ a unit vector normal to the surface of the sample 
\cite{Aharoni96}. Provided that the U and M states (as well as the vortex
wall) can be represented with $\mathbf{M}=M_{\phi }(z)\hat{\phi }+M_{z}(z)%
\hat{\mathbf{z}}$, it follows that the surface magnetic charges are given by 
$\sigma _{M}(z)=M_{z}(z)\hat{\mathbf{z}}\cdot \hat{\mathbf{n}}$, where $z=0$
or $L$. In the upper surface of the tube ($z=L$) $\hat{\mathbf{n}}=\hat{%
\mathbf{z}}$ and then $\sigma _{M}(L)=M_{z}(L)$, whereas in the lower
surface $\hat{\mathbf{n}}=-\hat{\mathbf{z}}$, and $\sigma _{M}(0)=-M_{z}(0)$%
. If the actual equilibrium state is the U or the M state, we find positive
magnetic charges ($\sigma _{M}(L)>0$) located at the upper surface, and this
region behaves like a source of charges, as in figure \ref{Ferro}; in
elementary electromagnetism the field lines emanate from positive charges.
Conversely, in the lower surface $\sigma _{M}(0)<0$ and the region behaves
like a sink of charge (the field lines converge to these points). It can be
shown that for nanotubes in the mixed state, the dipole field looks very
similar to the one plotted in figure \ref{Ferro}. As a consequence, we do
not present the same plots for tubes in the mixed state, instead we show in
figure \ref{Hdvsrho} the dependence of the magnitude of the dipole field ($%
H_{d}=\sqrt{H_{\rho }^{2}+H_{z}^{2}}$) with the radial coordinate for
different tube radius and $z=1050L_{x}$ (just $50L_{x}$ above the upper
surface of the tube). We have normalized all the curves to $%
M_{s}(R/L_{x})^{2}$, and in order to remark the difference between the U
state and the M state, we present groups of curves for both states. 

We observe that a proper consideration of the equilibrium state reduces
considerably the dipole field. This effect becomes critical as the tube
radius is increased, which increment the size of the vortex domains of the
mixed state \cite{LSC+09}. It is worth to mention about the importance of a
proper consideration of the actual magnetic state in tubular nanostructures.
In the sake of simplicity, it is a very common practice in magnetism to
approximate the magnetization as a uniform field, in spite that uniform
magnetization can be achieved (at zero applied field) only in nanostructures
with dimensions of the order of the exchange length \cite%
{Kravchuk,bookchapter}, or when the nanoparticle is an ellipsoid with the
appropriate size. Moreover, cylindrical magnetic nanowires and nanotubes are
frequently consider as infinite structures, and only in this limit the
approximation of a uniform dipole field becomes plausible, because the
cylindrical nanostructures can be considered as ellipsoids only if they are
infinitely long. However, actual magnetic nanostructures are not infinite
and finite size effects can be relevant. It is well known that the dipole
field produced by a nanoparticle with an ellipsoidal shape is uniform if its
size is below the single domain limit (in absence of an external field) \cite%
{Aharoni96}. In this limit one can safely write the dipole field in terms of
the corresponding demagnetizing factor: $\mathbf{H}_{d}=-N\mathbf{M}$. 
\begin{figure}[h]
\begin{center}
\includegraphics[width=7cm]{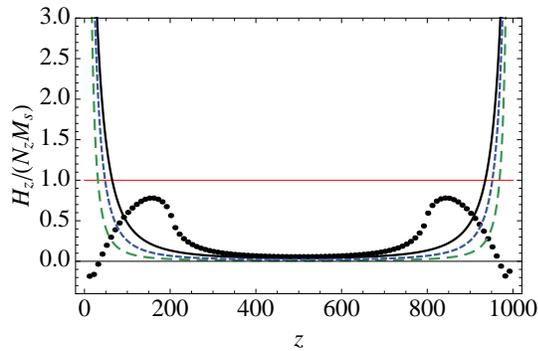}
\end{center}
\caption{(Colour online). Component of the dipole field along the MN axis as
a function of coordinate $z$. The fields have been normalized to $N_{z}M_{s}$%
, where $N_{z}$ is the corresponding demagnetizing factor. The long dashed
line corresponds to the U state for $R/L_{x}=4$. The short dashed line
corresponds to the U state and $R/L_{x}=10$. The thick solid line correspond
to U state and $R/L_{x}=20$ whereas the circles correspond to the M state
for $R/L_{x}=20$. The horizontal line illustrates the strength of the dipole
field calculated in the usual ellipsoid-like approximation for uniform
magnetization: $H_{d}=-N_{z}M_{s}$. }
\label{Nz}
\end{figure}

The difference between uniform and non-uniform dipolar fields is illustrated
in figure \ref{Nz}, where we have plotted $H_{z}$, the component of the
magnetostatic field along the tube axis, as a function of $z$ for different
tube radius and magnetic states. The horizontal solid line depicts the
strength of the (uniform) dipole field ($N_{z}M_{s}$) produced by an
infinite MN with uniform magnetization, whereas the long dash, short dash
and solid lines correspond to the field produced by a MN with a uniform
magnetization state with radius $R/L_{x}=4$, 10 and 20, respectively. The
circles correspond to the dipole field produced by a MN with a mixed state
rather the incorrect uniform magnetization state, and with $R/L_{x}=20$. The
fields have been normalized to $N_{z}M_{s}$, where $N_{z}=N_{z}(R,L,\beta)$
is the corresponding demagnetizing factor for a nanotube \cite%
{ELA+07-1,Beleggia06}.

\subsection{Vector plots for a vortex wall}

\begin{figure}[h]
\begin{center}
\includegraphics[width=8cm]{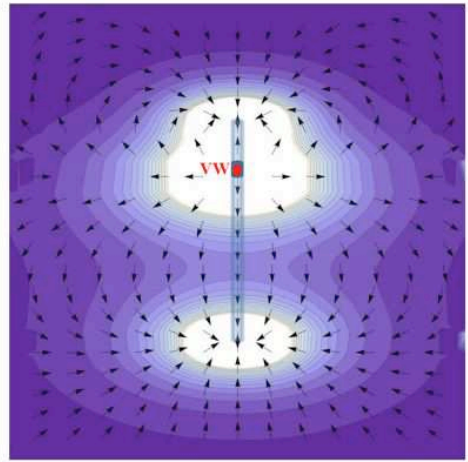}
\end{center}
\caption{(Colour online). Vector plot of the magnetostatic field for a
ferromagnetic nanotube with a vortex domain wall located at $z_{w}=750L_{x}$%
. The colours indicate the strength of the magnetostatic field. Parameters
are presented in the main text.}
\label{VW750}
\end{figure}

\begin{figure}[h]
\begin{center}
\includegraphics[width=8cm]{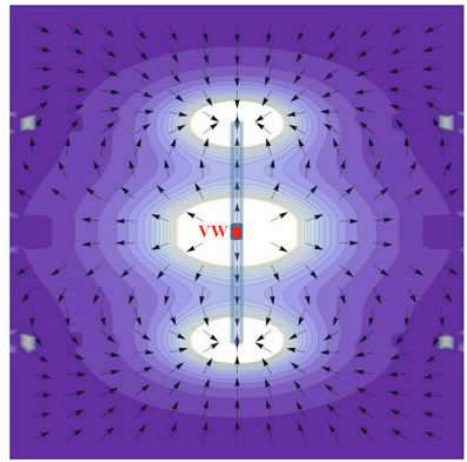}
\end{center}
\caption{(Colour online). Vector plot of the magnetostatic field for a
ferromagnetic nanotube with a vortex domain wall located at $z_{w}=500L_{x}$%
. The colours indicate the strength of the magnetostatic field. Parameters
are presented in the text.}
\label{VW500}
\end{figure}

In order to illustrate the basic characteristics of the magnetostatic field
produced by a MN with a vortex domain wall, we have chosen nanotubes with $%
\beta =0.9$, $L=1000L_{x}$ and $R=10L_{x}$. Following a previous work by
some of us \cite{LAE+07}, we can calculate numerically the DW width ($w$) by
minimizing the total energy including exchange. It is found that $%
w=w(\beta,R)$, and for the values mentioned above we obtain a wall width $%
w=109L_{x}$. In figure \ref{VW750} we show the magnetostatic field
normalized to his strength for a vortex wall located at $z_{w}=750L_{x}$,
whereas figure \ref{VW500} shows the same plot for the VW confined at the
middle of the tube ($z_{w}=500L_{x}$). 

Both figures illustrate three regions where there are magnetic charge; upper
and lower tube ends, and the region of the DW around the wall position $%
z_{w} $, which has been highlighted with red colour. There are negative
surface charges at the tubes ends, because $\sigma
_{M}(0)=-M_{z}(0)=-M_{s}<0 $ and $\sigma _{M}(L) =M_{z}(L)=-M_{s}<0$. Also,
there is a positive volume charge localized in the domain wall region. A
volume magnetic charge \cite{Aharoni96} is given by $\rho _{M}=-\nabla \cdot 
\mathbf{M}$ , and thus in our case $\rho _{M}=-\partial M_{z}/\partial z$.
Besides, $M_{z}=M_{s}\cos \Theta (z)$, and therefore, $\rho
_{M}=M_{s}(\pi/w)\sin \Theta>0$, just in the DW region. Therefore, the DW
acts as a source of magnetic charge. Note that we are considering here
head-to-head DWs, and in order to analyze the case of tail-to-tail walls, we
should replace $M_{z}$ by $-M_{z}$ and the corresponding magnetic charges
must have the opposite sign as discussed here for head-to-head walls.

\begin{figure}[h]
\begin{center}
\includegraphics[width=8cm]{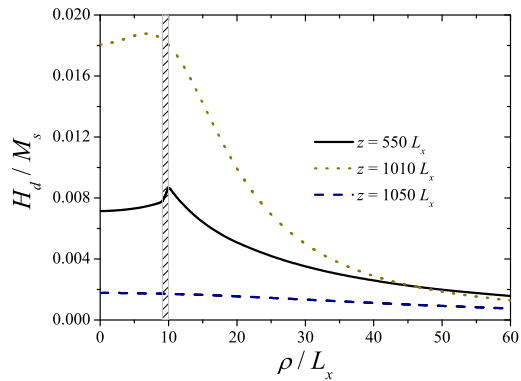}
\end{center}
\caption{(Colour online). Strength of the magnetostatic field for a MN with
a vortex domain wall located at $z_{w}/L_{x}=500$. Parameters are the same
of figure \protect\ref{VW500} while the dashed region represent the region
occupied by the magnetic matter. }
\label{Hdstrength}
\end{figure}

Finally, in figure \ref{Hdstrength} we illustrate the strength of the
dipolar field as a function of the distance to the z axis, and for different 
$z$ values. We have considered the same parameters as in figure \ref{VW500}
which means that we have magnetic matter just for $\rho /L_{x} = 9-10$ and $%
z/L_{x} = 0-1000$.

\subsection{Vector plots for a transverse wall}

To visualize the dipole field produced by a transverse wall, we have chosen
MNs with the same parameters of the above section, that is $\beta =0.9$, $%
L=1000L_{x}$ and $R=10L_{x}$. Following a previous work \cite{LAE+07} we can
calculate the transverse DW width ($w$) by minimizing the total energy, as
stated before. Again, it is found that $w=w(\beta,R)$, and for the values
mentioned above we obtain $w=7.67L_{x}$, a rather small value as compared to
the corresponding DW width for the VW of the above section. The magnetic
field for the transverse wall configuration is given by equations (\ref%
{HTrho}-\ref{HTz}). Note that the angular component of the field $H_{\phi }$
is not zero (unless $\sin \phi =0$) and therefore the field does not present
azimuthal symmetry as the magnetostatic field for the vortex DW. It can be
shown that for the geometry discussed here, the vector field along the plane
defined by $\phi =0$\ lies completely in that plane, where $H_{\phi }(\phi
=0)=0$. Along this plane it is found that the field is very similar to the
field for the VW (see figure \ref{VW500}). This can be understood by noting
that, mathematically, the field for the TW is very similar to the case of
the VW. The main differences are the wall width and the term dependent on
the angle $\phi $, and the function $A(\rho,z)$ defined in (\ref{A}).

\begin{figure}[h]
\begin{center}
\includegraphics[width=8cm]{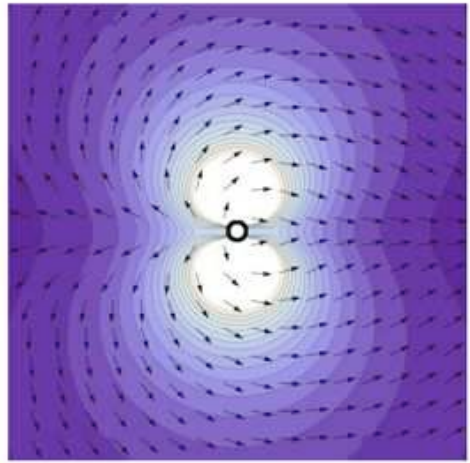}
\end{center}
\caption{(Colour online). Magnetostatic field for a ferromagnetic nanotube
with a transverse domain wall located at the middle of the tube ($%
z_{w}=500L_{x}$). The plot represent the top view of the field, at the plane 
$z=600L_{x}$. The colour indicates the strength of the magnetostatic field,
and along the contours the value of the field is constant. Parameters are
presented in the text.}
\label{TW1}
\end{figure}

\begin{figure}[h]
\begin{center}
\includegraphics[width=8cm]{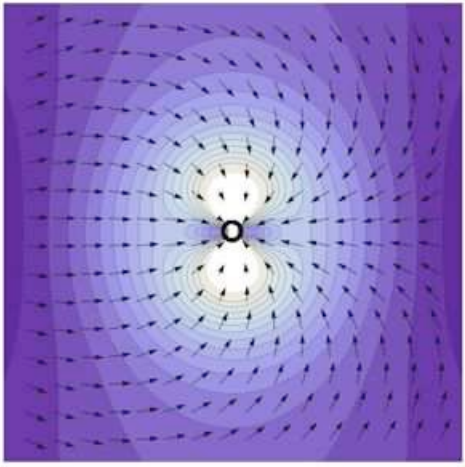}
\end{center}
\caption{(Colour online). Magnetostatic field for a ferromagnetic nanotube
with a transverse DW located at the middle of the tube. The plot represent
the top view of the field at the plane $z=1200L_{x}$. The colours indicate
the strength of the magnetostatic field. Along the contours the value of the
field is constant. Parameters are presented in the text.}
\label{TW2}
\end{figure}

In figures \ref{TW1} and \ref{TW2} we show the magnetostatic field produced
by the magnetization of a transverse DW located at $z_{w}=500L_{x}$. Figure %
\ref{TW1} depicts the top view of the field along the plane $z=600L_{x}$,
and figure \ref{TW2} shows the plane $z=1200L_{x}$. Both plots represents
the in-plane component of the field ($H_{\rho }^{TW}\hat{\rho }+H_{\phi
}^{TW}\hat{\phi }$) normalized to their strengths given by the colour code.
The central ring illustrates the top view of the nanotube. We have also
included information about $H_{z}^{TW}$ which is given by the length of the
arrows. It can be noted that near the ferromagnetic tube, the z component of
the field is more important, and thus the arrows are smaller.

\section{Final Remarks}

\label{Remarks}

In summary, we have investigated the dipole field produced by equilibrium
and non-equilibrium domain wall states of magnetic nanotubes. Using a
continuous model we have obtained simple expressions to evaluate the dipole
fields. On one hand, we can conclude that the consideration of magnetic
states, as the non-uniform mixed state instead the idealized uniform state,
reduces considerably the magnetostatic field. Thus, it is important for
researchers to have in mind this approach when they compare with
experimental results, in order to avoid the overestimation of stray fields.
On the other side, magnetic domain walls, which can be manipulated by
external fields or spin-currents, introduce volumetric magnetic charges into
the nanostructure. These charges move in the direction of the applied field
if the DW is a head-to-head wall, and in the opposite direction if the wall
is a tail-to-tail one. Finally, we have concluded that the magnetostatic
field strongly depends on the mechanisms of magnetization reversal. Our
results are intended to provide guidelines to use the magnetostatic field
generated by tubular nanostructures in prospective applications such as the
generation of a magnetic trap.

\section*{Acknowledgments}

This work was partially supported by FONDECYT Grants No. 11080246 and
11070010, Financiamiento Basal para Centros Cient\'{\i}ficos y Tecnol\'{o}%
gicos de Excelencia, Millennium Science Initiative under Project P06-022-F
and the program \textquotedblleft Bicentenario en Ciencia y Tecnolog\'{\i}%
a\textquotedblright\ PBCT under project PSD-031.


\begin{thebibliography}{99}
\bibitem{WAB+01} Wolf S A, Awschalom D D, Buhrman R A, Daughton J M, von
Molnar S, Roukes M L, Chtchelkanova A Y and Treger M 2001 \textit{Science} 
\textbf{294} 1488.

\bibitem{GBH+02} Gerrits Th, van den Berg H A M, Hohlfeld J, Bar L and
Rasing Th 2002 \textit{Nature} (\textit{London}) \textbf{418} 509.

\bibitem{ET03} Emerich D F and Thanos C G 2003 \textit{Expert Opin. Biol.
Ther.} \textbf{3} 655.

\bibitem{Lee07} Lee D, Cohen R E and Rubner M F 2007 \textit{Langmuir} 
\textbf{23} 123

\bibitem{Eisenstein05} Eisenstein M 2005 \textit{Nat. Methods} \textbf{2}
484.

\bibitem{SRH+05} Son S J, Reichel J, He B, Schushman M and Lee S B 2005 
\textit{J. Am. Chem. Soc.} \textbf{127} 7316.

\bibitem{NCM+05} Nielsch K, Castano F J, Matthias S, Lee W and Ross C A 2005 
\textit{Adv. Eng. Mater.} \textbf{7} 217.

\bibitem{WLL+05} Wang Z K, Lim H S, Liu H Y, Ng S C, Kuok M H, Tay L L,
Lockwood D J, Cottam M G, Hobbs K L, Larson P R, Keay J C, Lian G D and
Johnson M B 2005 \textit{Phys. Rev. Lett.} \textbf{94} 137208.

\bibitem{TGJ+06} Tao F, Guan M, Jiang Y, Zhu J, Xu Z and Xue Z 2006 \textit{%
Adv. Mater. (Weinheim, Ger.)} \textbf{18} 2161.

\bibitem{Hertel04} Hertel Riccardo, Kirschner J\"{u}rgen 2004 \textit{J.
Magn. Magn. Mater.} \textbf{278} L291.

\bibitem{DKG+07} Daub M, Knez M, Gosele U and Nielsch K 2007 \textit{J.
Appl. Phys.} \textbf{101} 09J111.

\bibitem{ELA+07-1} Escrig J, Landeros P, Altbir D, Vogel E E and Vargas P
2007 \textit{J. Magn. Magn. Mater.} \textbf{308} 233.

\bibitem{ELA+07-2} Escrig J, Landeros P, Altbir D and Vogel E E 2007 \textit{%
J. Magn. Magn. Mater.} \textbf{310} 2448.

\bibitem{LSC+09} Landeros P, Suarez O J, Cuchillo A and Vargas P 2009 
\textit{Phys. Rev. B} \textbf{79} 024404.

\bibitem{LSS+07} Lee Johyun, Suess Dieter, Schrefl Thomas, Hwan Oh Kyu and
Fidler Josef 2007 \textit{J. Magn. Magn. Mater.} \textbf{310} 2445.

\bibitem{CUB+07} Chen A P, Usov N A, Blanco J M and Gonzalez J 2007 \textit{%
J. Magn. Magn. Mater.} \textbf{316} e317.

\bibitem{AAX+03} Atkinson D, Allwood A, Xiong G, Cooke M D, Faulkner C C and
Cowburn R P 2003 \textit{Nat. Mater.} \textbf{2} 85.

\bibitem{THJ+06} Thomas L, Hayashi M, Jiang X, Moriya R, Retener C and
Parkin S S P 2006 \textit{Nature} (\textit{London}) \textbf{443} 197.

\bibitem{LAE+07} Landeros P, Allende S, Escrig J, Salcedo E, Altbir D and
Vogel E E 2007 \textit{Appl. Phys. Lett.} \textbf{90} 102501.

\bibitem{EBJ+08} Escrig J, Bachmann J, Jing J, Daub M, Altbir D and Nielsch
K 2008 \textit{Phys. Rev. B} \textbf{77} 214421.

\bibitem{AEA+08} Allende S, Escrig J, Altbir D, Salcedo E and Bahiana M 2008 
\textit{Eur. Phys. J. B} \textbf{66} 37.

\bibitem{BEP+09} Bachmann Julien, Escrig Juan, Pitzschel Kristina, Montero
Moreno Josep M, Jing Jing, Gorlitz Deflet, Altbir Dora and Nielsch Kornelius
2009 \textit{J. Appl. Phys.} \textbf{105} 07B521.

\bibitem{EAA+08} Escrig J, Allende S, Altbir D and Bahiana M 2008 \textit{%
Appl. Phys. Lett.} \textbf{93} 023101.

\bibitem{PDE09} Pereira A, Denardin J C and Escrig J 2009 \textit{J. Appl.
Phys.} \textbf{105} 083903.

\bibitem{EAA+09} Escrig J, Allende S, Altbir D, Bahiana M, Torrejon J,
Badini G and Vazquez M 2009 \textit{J. Appl. Phys.} \textbf{105} 023907.

\bibitem{KVL+03} Klaui M, Vaz C A F, Lopez-Diaz L and Bland J A C 2003 
\textit{J. Phys.: Condens. Matter }\textbf{15}, R985.

\bibitem{CRE+00} Casta\~{n}o F J, Ross C A, Eilez A, Jung W and Frandsen C
2000 \textit{Phys. Rev. B }\textbf{69}, 144421.

\bibitem{Aharoni96} Aharoni A 1996 \textit{Introduction to the Theory of
Ferromagnetism} (Clarendon Press: Oxford).

\bibitem{Kravchuk} Kravchuk Volodymyr P, Sheka Denis D and Gaididei Yuri D
2006 \textit{J. Magn. Magn. Mater.} \textbf{310} 116.

\bibitem{bookchapter} Landeros Pedro, Escrig Juan and Altbir Dora 2009 In 
\textit{Electromagnetic, Magnetostatic, and Exchange-interaction Vortices in
Confined Magnetic Structures} Edited by E. O. Kamenetskii (Research
Signpost: Kerala).

\bibitem{Beleggia06} Beleggia M, Lau J W, Schofield M A, Zhu Y, Tandon S and
De Graef M 2006 \textit{J. Magn. Magn. Mater.} \textbf{301} 131.
\end{thebibliography}
\end{document}